\documentclass[12pt, oneside]{article}   	
\usepackage{geometry}                		
\usepackage[dvipdfmx]{graphicx}				
\usepackage{amssymb}
\usepackage{color}
\usepackage{lineno}
\usepackage{braket}
\usepackage[normalem]{ulem} 
\usepackage{url}
\usepackage{comment}

\DeclareRobustCommand{\Erase}{\bgroup\markoverwith{\textcolor{red}{\rule[.5ex]{2pt}{0.4pt}}}\ULon}

\title{Japan's Strategy for Future Projects \\ in High Energy Physics}
\date{March 20th, 2022}
\author{Japan Association of High Energy Physicists (JAHEP) }

\begin{document}
\maketitle

\vspace{0.5cm}
\begin{center}
Editors: \\
M.~Endo, 
K.~Hamaguchi, 
M.~Ibe, 
T.~Ishibashi, 
A.~Ishikawa,
M.~Ishino, 
M.~Ishitsuka, 
S.~Kanemura, 
M.~Kuriki, 
T.~Mori\footnote{Contact: mori@icepp.s.u-tokyo.ac.jp},
S.~Moriyama, 
H.~Nanjo,
W.~Ootani, and 
Y.~Sato
\end{center}

\vspace{1cm}
\begin{center}
{\bf Submitted to the Proceedings of the US Community Study \\
on the Future of Particle Physics (Snowmass 2021)}
\end{center}

\newpage




\section*{Executive Summary}
The current strategy for future projects of the Japanese high energy physics community
remains as described in 
the Final Report of the Committee on Future Projects in High Energy Physics, published in 2017~\cite{report2017}. 
The Recommendation part of the Final Report is excerpted in the next page.

This document updates the Final Report by adding developments and advances that have occurred since 2017. It is summarized as follows:

\noindent
\begin{itemize}
\item An $e^+e^-$ Higgs Factory is the most important next large-scale particle physics facility. 
{\em The International Linear Collider (ILC), being the most advanced candidate, should be realized in Japan as early as possible}. 
Carry out the time-critical development work and support international discussions to start the preparatory phase in a timely manner. 
In parallel, continuing studies of new physics should be pursued using the LHC and its upgrades.

\item {\em Hyper-Kamiokande, currently under construction, should be completed as scheduled to study CP symmetry in the lepton sector as well as proton decays}. 
In the meantime search for CP violation should be continued with the T2K experiment and upgrades of the J-PARC neutrino beam.
\end{itemize}

The High Energy Physics Committee should pursue all available options to achieve the early realisation of these key, large-scale projects. \\

It is important to accumulate an increasing amount of data as planned for physics studies at SuperKEKB that started full operation in 2019. 
Some of the medium- and small-scale projects 
currently under consideration have implicit potential
to develop into important research fields in the future, 
as happened with neutrino physics.  
They should be promoted
in parallel in order to pursue new physics from various directions. 
Flavor physics experiments, such as muon experiments at J-PARC, 
searches for dark matter and neutrinoless double beta decay,
observations of CMB B-mode polarization and dark energy, 
are considered to be projects that have such potential.\\

Furthermore, accelerator R\&D should be continued 
to dramatically increase particle collision energies 
in preparation for future experimental efforts that 
may indicate the existence of new particles and 
new phenomena at higher energy scale.

\newpage
\begin{quotation}
\begin{sffamily}

\begin{center}
{\em Recommendation} (September 2017)~\cite{report2017}
\end{center}
\vspace{0.5cm}

In 2012, not only was a Higgs boson with a mass of 125 GeV discovered 
at the LHC, but three-generation neutrino mixing was also established.
Taking full advantage of the opportunities provided by these discoveries,
the committee makes the following recommendations concerning large-scale projects, 
which comprise the core of future high energy physics research in Japan.

\begin{itemize}
	\item 
        With the discovery of the 125 GeV Higgs boson at the LHC,
	{\em construction of the International Linear Collider (ILC) 
	with a collision energy of 250 GeV 
	should start in Japan immediately}
	without delay so as to guide the pursuit of particle physics 
	beyond the Standard Model 
	through detailed research of the Higgs particle.
	In parallel, continuing studies of new physics should 
	be pursued using the LHC and its upgrades.

	\item 
	Three-generation neutrino mixing has been established
        and has provided a path to study CP symmetry in the lepton sector.
        Therefore, {\em Japan should promote the early realization of Hyper-Kamiokande
	as an international project} due to its superior proton decay sensitivity,
	and should continue to search for CP violation with the T2K experiment 
	and upgrades of the J-PARC neutrino beam.
\end{itemize}

\noindent %
The High Energy Committee should pursue all available options 
to achieve the early realization of these key, large-scale projects.\\

It is important to complete the construction of SuperKEKB 
and start physics studies as scheduled.
Some of the medium- and small-scale projects 
currently under consideration have implicit potential
to develop into important research fields in the future, 
as happened with neutrino physics.  
They should be promoted
in parallel in order to pursue new physics from various directions. 
Flavor physics experiments, such as muon experiments at J-PARC, 
searches for dark matter and neutrinoless double beta decay,
observations of CMB B-mode polarization and dark energy, 
are considered to be projects that have such potential.\\

Furthermore, accelerator R\&D should be continued 
to dramatically increase particle collision energies 
in preparation for future experimental efforts that 
may indicate the existence of new particles and 
new phenomena at higher energy scale.


\end{sffamily}
\end{quotation}

\newpage

\tableofcontents
\newpage

\begin{center}
\begin{figure} 
	\includegraphics[width=14.5cm]{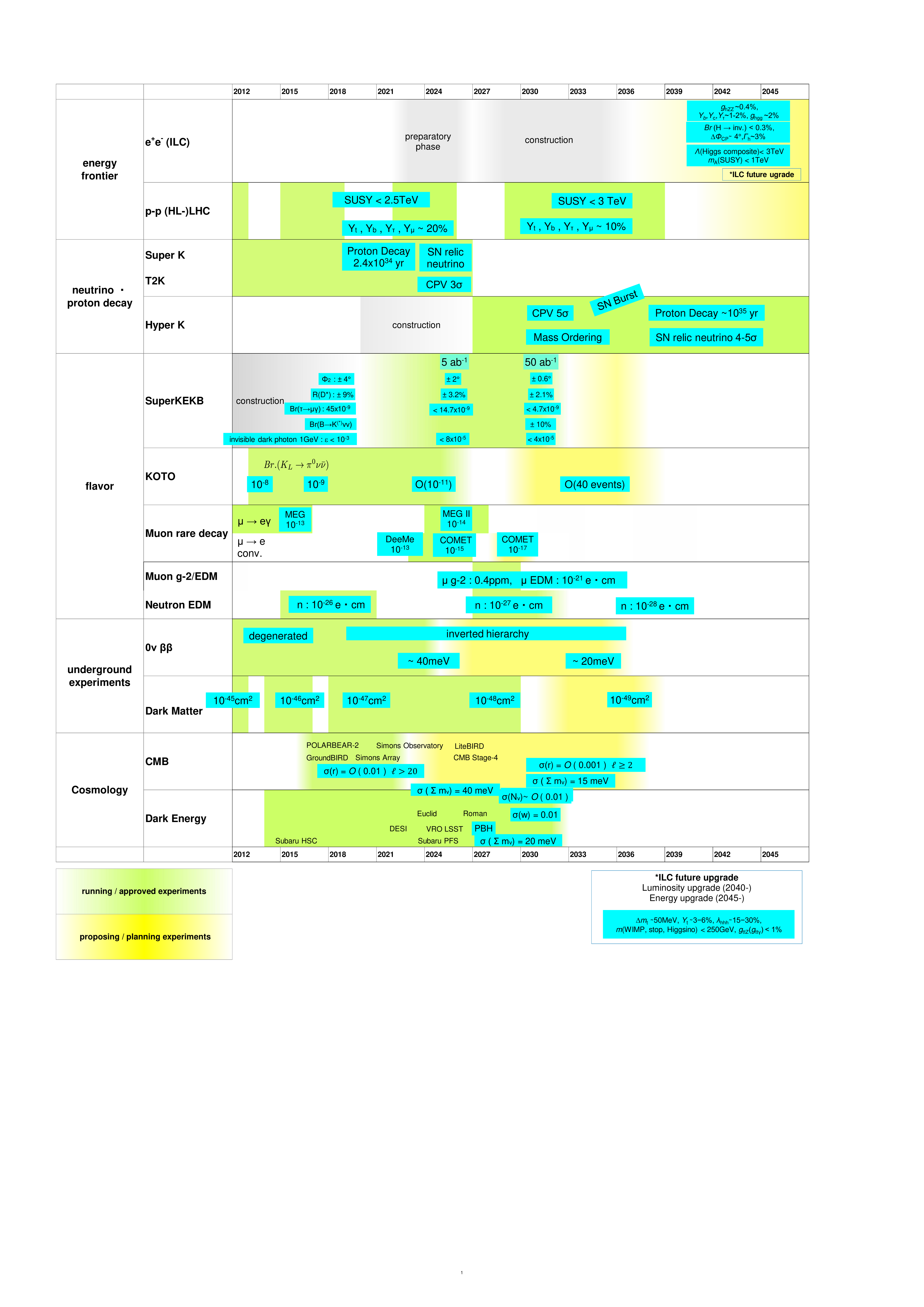}
	\caption{Timeline of ongoing and future projects described in the report.
	Sensitivities are overlaid at the time when they will be achieved (2012 -- 2040).}
	\label{fig:timeline}
\end{figure}
\end{center}

\section{Status and Prospects for Particle Physics}

The standard model of particle physics is the current best theory to describe the most basic building blocks of the universe.
It explains how all known matters are made up and how the electromagnetic, strong and weak forces act on them.
The theory has succeeded in explaining almost all experimental results and predicting a wide variety of phenomena very precisely.

In spite of these great successes, the standard model is known to describe physics processes
inadequately up to the Planck scale. 
With the discovery of the Higgs boson in 2012, experimental investigations of 
the Higgs sector at high energy accelerators have become the most solid path 
to approach more fundamental theories. 
To verify the mass generation mechanism of the standard model, it is really essential 
to make the best possible measurements of Higgs boson couplings to gauge bosons 
and quarks and leptons. 
Deviations from the standard model predictions, if experimentally confirmed, 
would determine the direction of new physics beyond the standard model 
through their patterns and magnitudes. 
To this end, precise measurements at the level of a percent or better are required, 
which can be realized only by lepton colliders such as 
the International Linear Collider (ILC).

The Higgs potential, the essential ingredient for the electroweak symmetry breaking, 
remains largely unknown and must be experimentally scrutinized. 
The Higgs potential does not only connect to the dynamics of the spontaneous symmetry 
breaking, but also determines aspects of the electroweak phase transition 
in the early Universe, which is critically important to understand 
the scenario of electroweak baryogenesis. 
The triple Higgs boson coupling, the most important parameter of the Higgs potential, 
can be measured at the high luminosity LHC (HL-LHC), but for sufficient precision 
the energy-upgraded ILC or next-generation high-energy hadron colliders would be 
absolutely necessary. 
The electroweak phase transition can be further explored through the synergy with 
future gravitational wave observations such as the LISA.  

Direct searches for new particles predicted by a variety of new physics models, 
such as supersymmetry or extra dimensions, should be performed at future 
energy frontier experiments. 
New particles, if discovered at LHC or HL-LHC, should be thoroughly studied at the ILC 
or the upgraded ILC to further deepen our understanding of the Universe. 
It will open a new era of particle physics. 

Another superior approach to probe higher energy scales, 
even beyond the kinematic reach of experiments, 
is to examine huge numbers of particles with sensitive detectors. 
It encompasses searches for extremely rare processes and for tiny deviations 
from standard model expectations induced by quantum effects.
Recently, the Fermilab Muon $g-2$ experiment announced a new measurement of 
the magnetic moment of the muon and confirmed the discrepancy found two decades earlier 
at Brookhaven National Laboratory, supporting possible existence of physics 
beyond the standard model. 
A striking pattern of anomalies, such as lepton-flavor universality violations, 
have also been emerging in rare $B$ meson decays, 
hinting at characteristic aspects of new physics.
Experimental explorations of these anomalies is therefore essential.
It is also important to search for other physical processes, 
such as flavor violations and electric dipole moments, 
as they give us clues to physics beyond the standard model.
Japan's two world-class accelerators, SuperKEKB and J-PARC, are among 
the best experimental facilities to study these key processes of flavor physics.

The discovery of neutrino oscillations established tiny masses of the neutrinos, 
which imply the existence of physics beyond the standard model.
The pattern of flavor mixing in the neutrino sector, being totally
different from that in the quark sector, 
may suggest that the origins of the neutrino and quark masses may be totally different.
Essential questions in neutrino physics include 
whether or not oscillations violate CP, 
whether neutrinos are Dirac or Majorana fermions, 
what are the absolute values of the neutrino masses, 
and whether the neutrino mass hierarchy is normal or inverted.
Precise measurements of CP violation and the determination of the neutrino mass hierarchy
will be carried out by future experiments, namely Hyper-Kamiokande in Japan 
and the DUNE experiment in the U.S.
Searches for neutrinoless double $\beta$ $(0\nu\beta\beta)$ decays are important 
to settle the issue of whether the neutrino mass is Majorana- or Dirac-type. 
If neutrinos are Majorana fermions, their masses could be related to physics 
at extremely high energy scales, as predicted by the seesaw mechanism. 
The Majorana nature and the CP violation in the neutrino sector would strongly imply that
the leptogenesis scenario might have given birth to the matter-antimatter asymmetry.

The search for proton decay, which is considered a signature of Grand Unification, 
has been explored by Super-Kamiokande with the world's highest sensitivity.
So far a proton decay signal has not been observed and the lower limit
on the proton lifetime has been established as $10^{34}$ years 
for the decay mode $p\rightarrow e^{+} + \pi^{0}$ and $p \rightarrow \bar{\nu} + K^{+}$.
Hyper-Kaiokande, DUNE and JUNO experiments will reach unprecedented sensitivities 
to the proton lifetime.

Connections between cosmology and particle physics have never been so strong.
Cosmological puzzles, such as the origin of the matter-antimatter asymmetry 
in the universe, the nature of the dark matter and dark energy and 
the cosmological inflation, seem to require some new ingredients in particle physics.
Naturally, any new ingredients beyond the standard model must have played 
important roles in the evolution of the universe. 
Running and future experiments will continue to provide indispensable knowledge 
and insights to solve these cosmological problems. 
For example, various experimental programs are being carried out or prepared 
to explore a broader range of dark matter candidates, 
including weakly interacting massive particles (WIMPs),
coherent oscillation of scalar condensation such as axions, 
and primordial black holes.
Weakly interacting dark matter particles can be detected by ongoing or future 
direct dark matter detection experiments, and their properties may be later 
investigated at high energy colliders such as LHC, HL-LHC and the ILC.
Observations of the cosmic microwave background (CMB), in particular, 
the B-mode polarization, will help deepen our understanding of 
the cosmological inflation and its energy scale. 
Progress in the study of dark energy and the origin of the accelerating expansion 
of the universe is also expected by upcoming programs to precisely measure 
the Hubble expansion rate and the angular distance over a wide redshift range.

\newpage

\section{Energy Frontier Physics}
\subsection*{Overview}


Research at energy frontier aims to discover new particles and phenomena as well as to determine their properties using state-of-the-art high-energy accelerators, and thereby to explore the laws of the nature governing them. 
Theories beyond the standard model are clearly necessary to explain dark matter, 
baryon asymmetry of the Universe and neutrino oscillation as well as 
the hierarchy problem. 
The proposed hypothetical ideas must be thoroughly tested by energy frontier experiments. 
A solid target is to precisely measure the Higgs boson couplings to other particles. 
The Higgs self-coupling must also be measured to explore the nature of the 
electroweak symmetry breaking and the phase transition at the early Universe. 
Direct searches for new particles and phenomena must be continuously pursued. 

In order to study higher energy phenomena at hadron colliders such as LHC, 
either the beam energy must be increased with stronger bending magnets 
to keep particles on their circular trajectory, 
or the luminosity must be increased to obtain a higher collision rate 
of high-energy partons.  
For lepton colliders, 
collision energy can be greatly extended at an electron-positron linear collider
without acceleration losses due to synchrotron radiation. 
Alternatively a circular muon collider may be developed.

At hadron colliders, the beams consist of protons or antiprotons, 
which are composite particles, so that a wide variety of phenomena 
can be studied from a broad perspective depending on the collision energies, 
including production and decays of the Higgs boson and searches for new particles. 
There are, however, issues from multiple interactions in the same beam bunch 
(pile-up) and large backgrounds. 
At lepton colliders, signals are clean because all the beam energy is converted 
into the reaction, so that the Higgs boson can be precisely studied. 
Once new particles are discovered, their properties can also be thoroughly measured. 
In particular, at a linear collider like the ILC, the beam energy and polarization 
can be adjusted to determine detailed properties of the new particles.

There is an international consensus for an electron-positron Higgs factory 
as the next highest-priority collider to be built. 
Several electron-positron collider projects are in progress including linear colliders
such as the ILC and the CLIC with future energy extendability beyond the Higgs factory, 
and circular colliders such as the FCC-ee/CEPC as Higgs Factories in the first step, 
followed by 100\,TeV-scale hadron colliders, FCC-hh/SppC in the second step.

\subsection{International Linear Collider}

While the importance of precision Higgs measurements is unquestionable, 
the physics case for higher collision energies has become increasingly strong. 
Among all the Higgs factory candidates, only linear colliders can extend
the collision energy to far beyond the Higgs factory.
Their capabilities to provide longitudinally polarized electron and positron beams 
are also great advantages in probing new physics. 
The ILC is currently the most technically mature and ready-to-be-built 
Higgs factory, which can be extended to much higher energies at later stages.
The 20~km machine with a collision energy of 250\,GeV can be 
extended to 30\,km (50\,km) with 500\,GeV (1\,TeV) 
with the currently available superconducting accelerator technologies. 
Recent rapid advances in accelerator technologies mean that 
even higher energy up to multi-TeV can be envisaged in the future.

The properties of the Higgs boson will be studied in great detail 
at the initial stage of the ILC with 250\,GeV. 
Many of the Higgs couplings to gauge bosons and fermions will be 
measured with a precision of 1\% or below,
which allows to pin down underlining new physics beyond the standard model. 
If the Higgs boson is a composite particle, its energy scale can be probed up to 3\,TeV.
If the Higgs boson is an elementary particle as in SUSY models, 
the mass of the heavy Higgs boson can be probed up to 1\,TeV.
The measurement of Higgs CP properties will shed light on baryogenesis models 
that can account for the asymmetry between matter and anti-matter in the universe.
At the upgraded stages of the ILC with 500\,GeV or above,  
the measurement of triple Higgs self-coupling will provide a direct probe
of the Higgs potential which will help reveal the deepest mystery in the
electroweak symmetry breaking. 

The ILC will also dramatically improve the precision of our understanding
of electroweak reactions. Already at the initial stage of the ILC the radiative 
return process $e^+e^-\to\gamma Z$ will provide 90 million $Z$-pole events.
Taking advantage of the beam polarizations, this process will improve 
the measurements of weak mixing angle ($\sin^2\theta_w$) as well as 
Left-Right Asymmetry ($A_{LR}$) by a factor of 10 over LEP and SLC. 
The measurements of anomalous triple gauge couplings and top-quark electroweak 
couplings can also be 
improved at the ILC by one order of magnitude over what was measured at LEP2 and 
the projections at the HL-LHC.

There is ample room for the ILC to directly discover SUSY particles with masses 
of up to half of the center-of-mass energies, being able to close up the loop-holes
in the models with nearly degenerate mass spectra of the Lightest Supersymmetric 
Particles (LSP) and the next to LSP. 
The latest experimental result on $(g-2)_\mu$ indicates larger deviation 
with the SM prediction. Light supersymmetric particles that can explain 
such deviations will be within the reach of the ILC.
Many models beyond the standard model require new gauge symmetries 
for which the clean di-fermion processes 
at the ILC will offer a great sensitivity, being able to directly produce new gauge 
bosons such as $Z^{'}$ with a mass up to the collision energy 
or to provide indirect sensitivity up to around 10\,TeV.

The ILC can also make available the most intense and highest-energy electron and 
positron beams for beam dump and dedicated fixed-target experiments 
that will, for example, study light weakly-interacting particles. 
This potential of the ILC is currently under active and intensive discussion 
among a broader scientific community.

A more detailed introduction of the physics potential at the ILC can be found in the
ILC White Paper for Snowmass 2021~\cite{ILC_Snowmass_WP}.

{\flushleft \underline{Recent Progress and Overall Plan}}\\

In October 2012, immediately after the Higgs discovery, 
JAHEP proposed a staging approach to start the ILC 
with 250\,GeV~\cite{JAHEP_ILC250_2012}. 
Later in July 2017, 
after observing that no new particle had been discovered at the early LHC, 
JAHEP made another proposal~\cite{JAHEP_ILC250_2017},
calling for the early realization of the ILC as a Higgs factory. 

In June 2020, the European Strategy for Particle Physics 
was updated~\cite{EuropeanStrategyforParticlePhysics}. 
It identifies an electron-positron Higgs factory 
as the highest-priority next collider.
It is stated that the timely realisation of the ILC in Japan would be 
compatible with the strategy and the European particle physics community 
would wish to collaborate.
The international consensus for an electron-positron Higgs factory 
as the next collider has strengthened the efforts 
towards the realization of the ILC in Japan 
with an increasing number of people from various sectors such as 
scientists, politicians, industry members, business circles, and local communities.

To advance the ILC project forward to the next step, 
the ILC International Development Team (IDT) was established 
by the International Committee for Future Accelerators (ICFA) 
in August 2020~\cite{ILC_IDT}.
The IDT, hosted by the KEK, is mandated to prepare 
the ILC Preparatory Laboratory (Pre-lab), 
based on MOUs among participating laboratories, as the first step. 
It is envisaged that, 
if the ILC is approved based on an inter-governmental agreement 
during the Pre-lab phase of about four years, 
the construction phase of about ten years will start by establishing the ILC Laboratory
among the partner countries/regions. 

To facilitate the transition to the preparatory phase, 
two new community organizations were established by JAHEP: 
the ILC Steering Panel in October 2020 and 
the ILC Japan in May 2021. 
The ILC Japan, in close cooperation with KEK, communicates and discusses 
matters related to the ILC with the Japanese government and other stakeholders 
to advance the ILC project on behalf of JAHEP, 
while the ILC Steering Panel leads bottom-up activities to promote the project, 
cooperating with other scientific communities, government authorities, industry, 
and regional partners~\cite{ILC_SteeringPanel}. 
    
The IDT completed the ILC Pre-lab proposal in June 2021, which describes 
the organizational framework and the work plan of the 
Pre-lab~\cite{InternationalLinearColliderInternationalDevelopmentTeam:2021guz}.
At the same time JAHEP and KEK summarized the key issues of the project in a report, 
and submitted it together with the ILC Pre-lab proposal 
to the Ministry of Education, Culture, Sports, Science and Technology (MEXT). 
In response MEXT started a second round of the MEXT Expert Panel 
in July 2021. 
The Panel's recommendations 
were published in February 2022~\cite{ILC_KEKNewsMEXTPanelReport2022}.

{\flushleft \underline{ILC Detector Technology R\&D}}\\
ILC detector technology R\&D has long been carried out internationally 
for two detector concepts; ILD and SiD.
Both are designed to achieve unprecedented 
performance of the hadron jet reconstruction with the particle flow approach, 
which requires high-precision pixel vertex detector, 
large-solid-angle tracking detector, and high-granularity calorimeter.
Japanese physicists play an important role in the global detector R\&D effort
including the international detector collaborations such as the LCTPC and the CALICE, 
with particular emphasis on the vertex detector, the TPC-based tracking detector, 
and the calorimeters for the ILD.
In close collaboration with theory groups, 
sensitivity studies and optimization of the detector concept are being advanced.
The Japanese detector operation experience with the Belle II, 
the ATLAS and the T2K near detector 
has provided useful information to develop important ILC detector components.

{\flushleft \underline{ILC Accelerator R\&D}}\\
Important technologies for ILC construction such as polarized electron
sources, positron beams, flat ultra-low emittance beams, high gradient
superconducting accelerating cavities, and nano-beam focusing are well 
established after more than 10 years of R\&D effort. The current aim of
ILC accelerator R\&D is further improvement of performance and
fabrication technology, detailed engineering design, and reliability studies 
by prototyping.

In recent years, various large scale superconducting accelerators have
been constructed and operated. 
The successful operations of these machines have demonstrated
the high performance and reliability of the ILC super-conducting
technology. At KEK-STF (Superconducting RF Test Facility), beam
acceleration with an average acceleration gradient of 33.0 MV/m,
exceeding the ILC specification, was achieved.

Hi-Q cavity with the nitrogen treatment and the flux expulsion had
been established. New treatment methods using oxygen diffusion which
is simpler than that with nitrogen demonstrated high-Q and high
accelerating gradient.

R\&D work for cost reduction has proceeded under the collaboration
between Japan and the United States. An example is cavity fabrication
with direct sliced Nb sheet leading to a cost reduction by omitting
several processes. Prototype cavities showed more than 35 MV/m.

The feasibility of the beam focusing at the interaction point of the ILC
has been demonstrated, by achieving a vertical beam size of 41 nm at
ATF2. Investigation for the emittance growth for a large bunch charge
at ATF2 showed that the effect is sufficiently suppressed at the ILC
with much higher beam energy and shorter bunch length. A study
to improve the luminosity is in progress by applying corrections to
higher-order components of magnet at ATF2.

Polarized electron source is technically established as the strain
compensated GaAs/GaAsP super-lattice (SCSL) cathode. The laser system
for spin polarized electron generation with a high stability and
reliability has to be designed.

Positron source development shows significant progress. For E-driven
positron source, a full-scale target prototype has been fabricated and
a long-term stable operation by keeping $7\times 10^{-7}$~Pa is confirmed. The
cooling design of the target, the flux concentrator, and the first
accelerator tube has been progressed, and the prototype will be
fabricated in 2022. The heavy beam loading at the capture linac is
compensated by a phase modulation on the input RF. In the Undulator
Positron source, the vitality of the target material (Ti alloy) was
examined with an electron beam.

{\flushleft \underline{Green ILC}}

Recently carbon neutral and sustainability are becoming an increasingly important
issue to consider. ``Green ILC'' is a series of initiatives to address
these issues and started in 2013. The expected $\rm CO_2$ emission from
the ILC is 320 kton/year, which would increase emission from the surrounding area 
(Ichinoseki city) 
by 24\%. 
Many studies on urban design, such as energy management with waste heat recycling 
and reuse and $\rm CO_2$ absorption by regional forests, 
and developments on new technologies, such as high efficiency klystron and
energy effective accelerator operation, have been 
conducted by the Green ILC Working Group (WG) of the Advanced Accelerator
Association (AAA), which is a tight collaboration among industry, academia,
and local governments. The WG contributes to the international panel on
sustainable colliders and accelerators that was established by ICFA in 2015.

\subsection{LHC Upgrade}

{\flushleft \underline{Introduction}}\\
Large-Hadron-Collider (LHC) has been running and producing physics results since 2010, and the center-of-mass energy reached 13\,TeV at the beginning of Run-2 in 2015. The discovery of the Higgs boson with a mass of 125\,GeV in 2012 is one of the outstanding results, and the properties, especially the coupling to the third generation fermions, has been confirmed and measured. The reach of new physics searches is extended by LHC, and the energy frontier as well as the luminosity frontier of the particle physics are being continuously developed.
The final focusing magnet of the LHC was designed and constructed by the collaboration of KEK (Japan) and Fermilab (USA) and it played a key role in achieving the instantaneous luminosity of 2.1$\times$10$^{34}$\,cm$^{-2}$s$^{-1}$ in Run-2, which is twice as high as the design value. The performance and stability of the final focusing magnet is one of the key factors for success.

The ATLAS Japan collaboration consists of 12 universities and KEK and has been making significant contributions to the design, construction, commissioning, and operation of various systems including the solenoid magnet, the silicon-strip detector, and the muon-trigger system. The grid computing system has become a vital tool for processing large-scale data and obtaining physics results. Japan has been making significant contributions as a Tier-2 center.
On top of these contributions, Japanese ATLAS collaborators are achieving outstanding physics results, examples of which are the discovery of the Higgs boson and continuous studies on the production and decaying processes. Further, important physics topics have been explored by young Japanese researchers and students, including contributions to searches for SUSY and new particles as well as precision measurements of the standard model particles.

{\flushleft \underline{Physics program in the LHC upgrade project}}\\
After the discovery of the Higgs boson, a new era in the exploration of physics beyond the standard model has begun.
Direct observations of new particles as well as precise measurements of the Higgs boson are effective approaches.
In terms of precision measurements of the Higgs boson, 
its couplings to the $W/Z$-boson, top quark, bottom quark, $\tau$ and $\mu$ will be measured
with a precision of 1 to 10\% with an integrated luminosity of 3000\,fb$^{-1}$.
It is crucial to test whether the couplings of the Higgs boson with other elementary particles are consistent with the predictions of the standard model
or a different rule is found. The differential cross-section of Higgs boson is a probe to explore new physics; especially the high-$p_{\rm{T}}$ tails are sensitive to new physics.
The measurement of the self-coupling of Higgs is also coming into view in the HL-LHC era.

A direct search for new particles is another main subject of the LHC upgrade program, and the strategy is to extend the reach by increasing the integrated luminosity significantly.
The leading target of the new physics search in the HL-LHC era is the SUSY particle observation, which is motivated by the presence of dark matter in the universe. In particular, mass degenerated Gauginos or Higgsinos are the primary focus, and the extension of the reaches within the parameter space which is consistent with the relic density of the dark matter is one of the critical goals. In addition, the naturalness still well motivates the third generation SUSY particle searches, although the LHC search results have excluded a part of the bulk parameter spaces. The high statistics HL-LHC data will allow us to explore the remaining phase spaces in the third generation SUSY searches.
Searches for other types of new particles will be continued and the reaches will be further extended by introducing innovative analysis techniques, trigger-level analysis, etc. To understand the vacuum stability, the precise measurement of the top-quark mass is an important subject. In case no new signal will be observed, a cross-section anomaly of the vector-boson scattering may tell us the energy scale to be explored with future accelerators.

{\flushleft \underline{Development of the High-Luminosity LHC}}\\
For the High-Luminosity LHC (HL-LHC) physics program, the accelerator complex will be upgraded and operated.
The major upgrades on the injectors have been performed during the Long-Shutdown 2 (2019 -- 2021), and the brightness of the proton bunch is increased by a factor of two.
During the Long-Shutdown 3 (2026 -- 2028), various major upgrades of the accelerator components are scheduled to kick off the High-Luminosity LHC project (HL-LHC) starting from the year 2029, including the replacement of the final focusing magnets, the beam-separation dipole magnets, and the upgrade of the beam collimators.
In terms of machine operation, the instantaneous luminosity is controlled to 7.5$\times$10$^{34}$cm$^{-2}$s$^{-1}$ by introducing the luminosity leveling technique ($\beta^{*}$-leveling) to maximize the integrated luminosity with longer beam lifetimes. The plan is to accumulate 3000 - 4000\,fb$^{-1}$ in about ten years. The HL-LHC project has been officially approved by the CERN council in June 2016.

The key components for the success of the HL-LHC project are the performance of the magnets placed around the beam collision points,
notably the final focusing quadrupole magnets and the beam separation dipole magnets.
To increase the luminosity significantly, the beam will be defocused before the collision point and then strongly squeezed ($\beta^{*}$\,=\,0.25\,m in 2018 will be changed to 0.15\,m).
To cope with the strong beam focusing, the bore size of the final focusing magnet needs to be enlarged and at the same time the coil peak field needs to be increased from 8.6 Tesla to 11.4 Tesla.
This will be realized with a new superconducting material, Nb$_{3}$Sn, which will be applied as an accelerator component for the first time. 
The final focusing quadrupoles magnets are produced by the collaboration of CERN and the U.S. DOE laboratories, while the beam separation dipole magnets, whose coil is made of NbTi, will be constructed by KEK.
After producing several 2-m long model magnets, a full-scale prototype of the beam separation dipole was produced in 2021 and the performance was tested. The nominal operational field of 5.6\,Tesla was confirmed with a reasonable margin and is ready for series production.

{\flushleft \underline{Detector developments for HL-LHC }}\\
There are two major upgrade tasks for the ATLAS spectrometer, one is to replace all the inner trackers and the other is to revise all the trigger and DAQ systems.
Aiming to keep the high tracking and vertexing performance in the high radiation background of HL-LHC, five layers of pixel detectors and four layers of strip detectors with highly radiation tolerant silicon sensors will be installed.
The Japanese group has experience in assembling the ATLAS silicon detector and there is a major silicon sensor vendor, Hamamatsu (HPK), in Japan. Taking these advantages, Japanese institutes have made strong contributions to the R\&D program of the pixel and strip sensors. The production of strip sensor is progressing in HPK, and is about to start for the pixel sensor. Further contributions to the pixel module assembly, installation, commissioning, operation are foreseen.

Concerning the triggering and DAQ, in order to record the target physics objects efficiently under the high luminosity condition, the first stage hardware trigger is designed with a longer latency of 10\,$\mu$s  as well as the increased acceptance rate of 1\,MHz. The numbers of the current system are 2.5\,$\mu$s and 100\,kHz, respectively.
Thanks to the longer latency, sophisticated trigger algorithms, that are usually used in offline analyses, can be introduced to the first stage trigger. The Japanese group has been working on the muon hardware trigger electronics. The system design is completed, and the production of several types of electronics modules and their integration tests are progressing. Developments of firmware and software are done in parallel and the team is in full swing to be ready for the system installation at the beginning of LS3 (2026).

{\flushleft \underline{Towards future energy frontier of hadron collider}}\\
Technology developments for the HL-LHC project may open the door to higher beam energy for future hadron colliders.
There is a plan to newly excavate a large-scale tunnel and increase the center-of-mass energy to 100\,TeV, Future Circular Collider, FCC-hh. A few Japanese institutes join the FCC collaboration and are working on the physics performance as well as the spectrometer design. 
The key technology for realizing the project is the high field accelerator magnets generating 16\,Tesla, which is twice as the current LHC magnet, or even higher field. 
A consensus can be found in the report of European Strategy 2020 that the highest priority is to identify the next technology including the aspects of performance, stability, cost for mass-production.

The R\&D program of the superconducting material of Nb$_3$Sn is progressing at the cryogenic science center in KEK collaborating with Japanese universities and companies with supports of CERN.
\subsection{Future Accelerators and Novel Acceleration Techniques}
{\flushleft \underline{X-band High-Gradient Accelerating-Structure }}\\
The Compact Linear Collider (CLIC) is a potential future linear collider which aims
to collide electrons and positrons at 3 TeV at the final stage of the project. 
It was proposed and developed as an international collaboration between CERN 
and other laboratories including KEK. 
The total length of CLIC is 11.4 km at the first stage with 380 GeV and 
50 km at the final state with 3 TeV. 
The acceleration frequency and gradient of the main linac are 12 GHz (X-band) 
and 72~MV/m (380~GeV) / 100~MV/m (3~TeV), respectively, 
in order to realize the compact size. 
It should be noted that the first 380 GeV stage can be realized 
with X-band klystrons feeding the high-power microwave to the main linac 
instead of the two-beam acceleration. 
The key technologies were demonstrated at the CLIC Test Facility and 
collaborating laboratories.

The high-energy community in the US has proposed a new concept of a 
normal-conducting electron-positron linear collider, 
named Cool Copper Collider ($\rm{C^3}$), 
with a compact footprint of 8 km based on the following two innovative technologies: 
(1) a distributed coupling accelerating structure, 
in which RF power is distributed to individual cells through a waveguide manifold, 
allowing a high degree of freedom in accelerating-structure design; 
(2) cryogenic-temperature operation of copper structures, 
which enables higher acceleration efficiency and higher-field operation, simultaneously.
The basic design is completed for a center-of-mass energy of 250 GeV 
(first stage, 70 MV/m) and 550 GeV (second stage, 120 MV/m) 
with a total length of 8 km. 
The key technologies need to be demonstrated at a demonstration facility 
in the near future.

For future normal-conducting linear colliders, X-band (11.4 GHz) accelerating-structure
development has been conducted by KEK in collaboration with SLAC and CERN for many years.
X-band high-gradient accelerating tubes can be fabricated by KEK and high-power tested 
at Nextef (New X-band Test Facility) in KEK. 
The recent focus is to develop this mature technology for various applications, 
e.g. for medical/industrial use.

{\flushleft \underline{Muon Collider}}\\
A muon collider is another candidate for a future energy-frontier lepton collider.
The muon projects at J-PARC developed essential technologies that would be 
useful for a muon collider in the MAP scheme~\cite{MuonC_MAP:2011map}.
The technology of high-intensity positron beam developed for the ILC was investigated 
as a way to realize a muon collider in the LEMMA scheme~\cite{MuonC_LEMMA:2018map}.
Also, a new scheme of muon collider that would use 
the low-emittance positive muon beam~\cite{ReAccUColdMu:2018ucm} 
discussed in Section 4.3 and
negative muon cooling utilizing the reaction of muon-catalyzed fusion
is under discussion. 

{\flushleft \underline{Laser-plasma accelerator, novel accelerator technologies}}\\
A laser-plasma accelerator has a potential to realize much higher 
gradients than conventional technology.
Recently high-quality beams with percent-level energy spread and 
micrometer-level emittance have been obtained. 
It has been pointed out that the increase in emittance in plasma acceleration is 
due to a mismatch of the betatron with the strong focusing field 
in the plasma at the injection and extraction. 
The mismatch can be eliminated and the emittance growth can be suppressed 
by applying a gradient to the plasma density.
Thus design studies of laser-acceleration-based practical machines have been started.

In recent years, research on thin-film superconducting cavities 
using high Tc superconductor such as Nb$_3$Sn and MgB$_2$ 
have been actively conducted all over the world 
to realize performances that exceed Nb cavities. 
The Nb$_3$Sn cavity, in particular, was demonstrated to achieve 
accelerating gradient of larger than 20 MV/m 
with Q value of higher than $1\times 10^{10}$ at 4.2 K. 
It is expected that a future Nb$_3$Sn accelerator could be operated by cryo-coolers, 
without liquid helium. 
Also recently an increasing number of investigations on superconducting cavities 
with multi-layered thin film structure have been carried out. 
It is a very promising technology that can exceed Nb cavities, 
as high accelerating gradient of around 100 MV/m is theoretically predicted. 
 

\newpage

\section{Neutrino Oscillation and Proton Decay}
\subsection*{Overview}
The Super-Kamiokande and T2K experiments have been producing world-leading results 
in neutrino physics.
As these projects continue the research, the successor experiment, Hyper-Kamiokande~\cite{Hyper-Kamiokande:2018ofw}, is currently being constructed 
by the international collaboration to start operation in 2027.
The Hyper-Kamiokande project is hosted by the University of Tokyo and KEK in Japan.

\subsection{Status}
{\flushleft \underline{Neutrino Oscillation}}\\
Three flavor mixing is now established while the existence of CP violation 
in the neutrino sector has been indicated by T2K.
Measurements of mixing and CP violation in the lepton sector will provide insight 
into the underlying symmetry mechanisms in comparison with that of quark sector.
In addition, if the unitarity is violated in neutrino mixing, it will be taken as a hint of new physics at an ultra-high energy scale.
In addition, the sign of the mass squared difference $\Delta m^{2}_{32} = m^2_3 - m^2_2$ is still unknown.
The ordering of the neutrino masses is an important ingredient to understand the origin of extremely small neutrino masses in comparison with the other fermions.\\

{\flushleft \underline{Proton Decay}}\\
Searches for proton decay are currently dominated by Super-Kamiokande (SK), which has placed 90\% C.L. limits on the proton lifetime as $2.4 \times 10^{34}$ years for $p \rightarrow e^{+} + \pi^{0}$ and $8.2 \times 10^{33}$ years for $p \rightarrow \bar{\nu} + K^{+}$.
Some GUT models predict $p \rightarrow e^{+} + \pi^{0}$ as the dominant decay mode, while $p \rightarrow \bar{\nu} + K^{+}$ is considered as a major decay mode in supersymmetric GUT models.
These models predict proton lifetimes at around $10^{34}$ to $10^{35}$ years, which is a target of the future projects to search for.

\subsection{Future Plans}
The T2K experiment has reported a hint of CP violation from the measurement of neutrino and antineutrino oscillation probabilities using the J-PARC neutrino beam~\cite{T2K:2019bcf}.
Therefore, it is necessary to continue the search for CP violation by T2K to confirm the existence of CP violation with higher statistics, and in parallel, complete the upgrade of the J-PARC neutrino beam and construction of the HK detector to achieve the precision measurement.

Hyper-Kamiokande (HK) is currently under construction to study neutrino CP violation over a wide range of $\delta_{CP}$ with the combination of an upgraded J-PARC neutrino beam.
The HK long-baseline project can accumulate more than 20 times larger statistics than the T2K experiment with a larger fiducial volume of the HK detector than that of the SK detector, and upgrade of J-PARC accelerator beam power.
This is a tremendous improvement as the current sensitivity of T2K is limited by the statistical uncertainties.

Sensitivity to CP violation is enhanced once the mass ordering is determined.
The strategy of HK is to determine the mass ordering by the observation of atmospheric neutrinos.
Future projects, such as JUNO and DUNE, have sensitivity to determine the mass ordering with different approaches from HK using reactor neutrinos and neutrino beam, respectively.
Therefore, these future projects are complementary to HK.
DUNE has the sensitivity to CP violation as well.

Construction of the HK detector has started in 2020 and the operation will begin in 2027.
HK can extend the search for proton decay by an order of magnitude and thereby has a good opportunity to provide direct evidence of the GUT models.

As for the sterile neutrino search, JSNS$^{2}$~\cite{JSNS2:2021hyk} has started data taking in 2020 at MLF in J-PARC, aiming to test the oscillation parameter regions suggested by LSND and MiniBooNE.
Investigation of Majorana properties is another important subject of neutrino physics.
Future projects for neutrinoless double beta decay ($0\nu\beta\beta$) search will be explained in Section~\ref{sec:particle_underground}.


{\flushleft \underline{J-PARC Accelerator and Neutrino Beam}}\\
The neutrino beam at J-PARC is produced from 30-GeV protons accelerated by its Main Ring (MR). 
By 2016 the beam power reached $\sim$ 500\,kW.
The J-PARC accelerator upgrade was budgeted and is progressing toward a beam power of 1.3\,MW for the MR, which exceeds its original design value of 750\,kW.
In 2022 the repetition rate will be increased by almost a factor of two from the current value of 2.5\,sec with an upgrade of the power supplies for the main magnets and the introduction of higher gradient RF cavities.
These upgrades will result in a beam power greater than 750\,kW by 2023.
Further upgrades of the RF power supplies and the beam monitors are scheduled by 2025.
The proton beam will be further optimized and stabilized with these upgrades, and the number of protons per spill will be increased. The beam power is expected to reach 1.3\,MW by around 2028.
The J-PARC neutrino beam facility upgrade is also progressing to operate with the 1.3-MW beam power in the high repetition rate.

For the long-term future, further upgrades towards the realization beyond 1.3-MW accelerator are being considered.
One possibility is to construct a new 8-GeV booster ring between the current Rapid Cycling Synchrotron (RCS) and MR to allow more protons to be injected with smaller emittance.
The design of the booster ring is under development in view of location and cost.
As a possible operational improvement, the parallel operation of the fast extraction beam for neutrino experiments and slow extraction beam for hadron experiments can be realized if a stretcher ring for the slow extraction is newly constructed as an addition to the current MR accelerator.
The stretcher ring can be constructed inside the same tunnel of the MR.
There is also an idea to arrange linac accelerators in the KEKB tunnel to make a 10-MW-scale proton driver after the completion of the SuperKEKB project.
The driver front-end to 1.2 GeV was designed in detail, including the electromagnetic design of half wave resonator at the front-end. The challenges for high duty target horn to receive the 10-MW beam are under discussion.

{\flushleft \underline{Hyper-Kamiokande}}\\
Hyper-Kamiokande (HK) is a next-generation large-scale water Cherenkov detector based on the technology established in SK with the upgrades of the detector scale and performance.
The detector is designed as a cylindrical structure with 68\,m diameter and 71\,m height, with a fiducial volume of about 188\,kiloton, which is approximately 8.4 times larger than SK.
The physics capabilities of HK cover a broad range of topics as summarized in this section.

HK will be able to precisely measure neutrino oscillation parameters using the J-PARC neutrino beam with extremely high sensitivity to CP violation.
Statistical uncertainty of $\nu_{\mu} \rightarrow \nu_{e}$ and $\overline{\nu}_{\mu} \rightarrow \overline{\nu}_{e}$ appearance signals will be 5\% (3\%) in 5 (10) years operation.
In order to suppress the systematic uncertainties in the measurements and further enhance the sensitivity to CP violation, upgrades of the T2K near detectors and a construction of new 1\,kton scale water Cherenkov detector (IWCD) are planned in HK.
As a supplementary project, test beam experiment (WCTE) is proposed at CERN to be conducted in 2023 with a 50\,ton scale water Cherenkov detector to validate the performance of small water Cherenkov detector implemented with multi-PMT modules which consist of multiple small PMTs encapsulated in a vessel.
Accumulating data for 10 years with the 1.3MW J-PARC neutrino beam, the $\delta_{CP}$ measurement precision is estimated to be better than $23^{\circ}$ ($7^{\circ}$ for $\delta_{CP}=0^{\circ}$).

Determination of neutrino mass ordering is one of the major physics targets of HK, not only because it may give hints toward the origin of the extremely light neutrino mass, but also it is also related to studies of Majorana neutrinos being explored by the neutrinoless double beta decay searches.
Although the sensitivity depends on the value of mixing angles, HK will be able to determine the mass hierarchy at a significance of 4--6$\sigma$ with observations of atmospheric neutrinos and J-PARC beam neutrinos in 10 years of operations.
It should be noted that the sensitivity to CP violation is enhanced once the mass ordering is determined by HK or the other experiments due to the degeneracy of the parameters.

HK also has an unprecedented sensitivity to proton decays.
In addition to the increased fiducial volume, atmospheric neutrino backgrounds will be largely suppressed by the improved performance of the detector with new photosensor.
The $3 \sigma$ discovery potential reaches to $1\times10^{35}$ years for the $p \rightarrow e^{+} + \pi^{0}$ and $3 \times 10^{34}$ for $p \rightarrow \bar{\nu} + K^{+}$ decay modes,  approximately one order better than the current limits, and can therefore test various GUT models.

Because of its huge fiducial volume, HK's observations of astrophysical neutrinos, such as solar neutrinos, supernova neutrinos, as well as searches for dark matter can lead to new discoveries.
In terms of the supernova neutrino observations, HK has an outstanding capability to observe neutrinos from supernova bursts within the galaxy or its near surroundings.
SK has started operation with Gd loaded water in 2020 aiming to discover as-yet unmeasured supernova relic neutrinos, and HK is expected to observe them with even higher statistics.
As the flux of the supernova relic neutrinos depends on the frequency of supernova bursts in the early universe, its observation may advance the study of the evolution of the universe.

These physics capabilities can be further enhanced by the international contribution such as additional PMTs.

\newpage

\section{Flavor Physics}
\subsection*{Overview}
Flavor-physics experiments can probe physics at high mass scales, even beyond those kinematic reaches. 
Currently, several experiments have reported hints of phyiscs beyond the standard model as in measurements of the muon anomalous magnetic moment ($g-2$) and 
a variety of measurements at the B-factories and the LHCb experiments. 
It is highly important to further test these signals, and a wide program of flavor measurements are also required to shed light on physics beyond the standard model because they are complementary to other searches.
Japan is playing an essential role in the field of flavor experiments, 
as represented by SuperKEKB/Belle~II, KOTO, MEG II, COMET, J-PARC muon g-2/EDM,  
TUCAN, neutron lifetime at J-PARC, and NOPTREX.

\subsection{SuperKEKB / Belle~II Experiment}

Among the results by KEKB/Belle, several anomalies are seen and could be interpreted as new physics 
in high energy scale, for example, lepton flavor universality~(LFU) in 
$B \to D^{(*)} \tau \nu$~\cite{Belle:2015qfa,Belle:2016dyj,Belle:2019gij} and 
isospin sum rule in $B \to K \pi$~\cite{Belle:2012dmz}.
LHCb experiment at CERN also found an anomaly of the LFU in $B^+ \to K^+ \ell^+ \ell^-$ 
in 2021~\cite{LHCb:2021trn}. 
To exploit the anomalies and new physics in flavor physics, the SuperKEKB/Belle~II project had been started. 
\\

SuperKEKB/Belle~II is the only operational $e^+e^-$ B-factory. It has a rich, unique and diverse physics program~\cite{Belle-II:2018jsg, Belle-II:2022zzz2}. 
{\flushleft \underline{Flavor Physics}}\\
The primary aim of the Belle~II experiment is to make new physics discoveries at a high-energy scale through flavor physics with $B$, $D$ mesons and $\tau$ leptons with high precision.
A test of unitarity triangle can probe new physics in $B^0-\bar{B}^0$ mixing. 
Only Belle~II can provide the precise measurements of all six observables for the triangle (e.g., three angles and three sides) to test the Kobayashi-Maskawa theory with an order of 1~\% precision which probe new physics scale up to 2000~TeV.
New sources of $CP$ violation in quark and lepton sectors might explain the baryon asymmetry of the universe. 
Belle~II exploited various types of $CP$ violation, such as time dependent and integrated $CP$ asymmetries, triple product correlation, and electric dipole moment using $B$, $D$ and $\tau$ decays.
Extensions of Higgs sector, which are required in many new physics models, predict the existence of charged Higgs bosons. 
The decays of heavy flavor quarks and leptons, such as inclusive $B \to X_s \gamma$, $B^+ \to \tau^+ \nu$ and
$\tau^- \to K_S^0 \pi^- \nu_{\tau}$ decays, are sensitive to the contributions of charged Higgs bosons.
The tests of LFU in $b \to c \tau \nu$ and $b \to s \ell^+ \ell^-$ are one of the central issues in $B$ physics. 
If the current central values of $R(D^{(*)})$ and $R_{K^{(*)}}$ are unchanged, 5$\sigma$ observations are possible around middle of 2020s. 
Further, angular analyses can constrain the Lorentz structure of the new physics.
The yet unobserved decays $b \to s \nu \bar{\nu}$ can give another inputs to the new physics which explains anomalies in $b \to s \ell^+ \ell^-$. 
Belle~II can observe the $B \to K^{(*)} \nu \bar{\nu}$ with 10\% precision.
Studies of charged lepton flavor violation and electric dipole moment of $\tau$ lepton are only possible at Belle~II among the running experiments. If these signatures are seen, it is a clear new physics signal. 

{\flushleft \underline{Dark Sector and Hadron Physics}}\\
Measurements of radiative return processes especially in $e^+e^- \to \pi^+\pi^-\gamma$ can improve the theoretical uncertainty of hadronic vacuum polarization contributions to muon $g-2$ in the SM. Belle~II will perform the measurement of $e^+e^- \to \pi^+\pi^-\gamma$ with similar or better sensitivity as that for BaBar.
The dark matter and its portal particle, which is light around MeV$\sim$GeV mass scale and has smaller coupling to the SM particles, can be searched for at clean and high luminosity $e^+e^-$ collider environment. 
One of the flagship search signatures at Belle~II is the $e^+e^- \to \gamma + {\rm nothing}$ which is predicted in many new physics models.
The mechanism of hadron formation will be studied which will lead to advances in elementary particle physics, hadron physics, astrophysics and their associated interdisciplinary fields.

{\flushleft \underline{Apparatus}}\\
SuperKEKB is an asymmetric-energy electron-positron double-ring collider with a 3-km circumference that aims to achieve a peak luminosity of $6 \times 10^{35} \, \mathrm{cm}^{-2} \mathrm{s}^{-1}$. 
The SuperKEKB design is based on the “nanobeam” scheme, wherein the beam size is reduced to 50-60 nm in the vertical direction and 10 $\mu$m in the horizontal direction at the interaction point (IP). 
Belle~II is a multi-purpose 4$\pi$ detector aiming to collect 50~$\mathrm{ab}^{-1}$ of data to provide the world's highest sensitivities for searches for new phenomena by around 2031.
The Belle~II consists of several sub-detectors to measure the decay vertices of $B$ mesons, and energy of charged and neutral particles precisely, and to identify charged particle flavor efficiently. 
The clean environment at $e^+e^-$ collider allows reconstruction of neutral particles (e.g., photon, $\pi^0$, $K_S^0$, and $K_L^0$) efficiently, which is crucial for inclusive measurements and tagging of the other $B$ mesons. 

{\flushleft \underline{Operation}}\\
The operation of SuperKEKB Phase 3 began in March 2019 with the Belle~II detector. 
The crab waist scheme was implemented to improve the beam-beam performance in April 2019. 
In 2020, SuperKEKB surpassed the luminosity record of $2.11 \times 10^{34} \, \mathrm{cm}^{-2} \mathrm{s}^{-1}$
achieved by its predecessor KEKB with the vertical and horizontal beta functions at the IP set at 1 mm 
and 80(LER)/60(HER) mm, respectively. 
Owing to the innovative nanobeam scheme, the peak luminosity record was surpassed with smaller beam currents 
than were used at KEKB. 
In December 2021, SuperKEKB set a new world record for a peak luminosity of $3.81 \times 10^{34} \, \mathrm{cm}^{-2} \mathrm{s}^{-1}$. 
Since 2019, Belle~II has collected 268~${fb}^{-1}$ of integrated luminosity and published world-leading physics results such as dark sector, $B \to K \nu \bar{\nu}$ and charm lifetimes~\cite{Belle-II:2019qfb,Belle-II:2020jti,Belle-II:2021cxx,Belle-II:2021rof}.

{\flushleft \underline{Upgrades}}\\
While achieving steady progress, SuperKEKB encounters some challenges as a luminosity frontier machine, such as the efficiencies and stabilities of the beam injection to the main ring, collimation and machine protection strategies, safe increase of beam currents, lowering the beta functions at the IP, and overcoming various beam instabilities.  
An International Task Force has been organized and an active exchange of ideas and opinions has been conducted in order to solve problems that hinder luminosity improvement.
SuperKEKB continues to improve its performances in the framework of the SuperKEKB/Belle-II collaboration and with the accelerator community in the world.

The Belle~II detector upgrades are planned to improve its performance significantly even under higher beam-induced background during long shutdown 1~(LS1) from 2022 summer and LS2 around 2026. 
The upgrade in LS1 is being prepared to install the second layer of DEPFET pixel detector and to replace the conventional MCP-PMTs in TOP to life-extended ones if performance requires. 
The major detector upgrade is proposed in LS2 to replace almost all sub-detectors and 
add new particle identification detector~\cite{Forti:2022mti}. 
The worldwide R\&D effort for the major upgrade is ongoing and EOI's had been submitted to the upgrade advisory committee in 2021.

Consideration of further luminosity upgrade and electron polarization capability of SuperKEKB are started for ultimate new physics searches with heavy flavor quarks and leptons including $\tau$ lepton $g-2$ in the light of muon $g-2$ anomaly~\cite{Crivellin:2021spu}.

\subsection{Kaon}
Measurements of the branching ratios of the decays
$K_{L} \rightarrow \pi^{0}\nu\overline{\nu}$ and
$K^{+} \rightarrow \pi^{+}\nu\overline{\nu}$
are among the most important
in the search for new physics with rare kaon decays.
The branching ratios are sensitive to new physics,
since those are extremely small
and precisely predicted in the standard model.

The NA62 experiment at CERN SPS measured the branching ratio  of $K^{+} \rightarrow \pi^{+}\nu\overline{\nu}$
with the data collected in 2016-18
to be $(10.6^{+4.0}_{-3.4}\pm 0.9)\times 10^{-11}$,
which is consistent with
the standard model branching ratio $8\times 10^{-11}$.
NA62 will measure it with a precision of $\sim 10\%$ in the 2020s.
After that, the NA62 group is considering comprehensive programs to study rare decay modes for both $K^+$ and $K_L$,
including measurements of the branching ratios of
$K^{+} \rightarrow \pi^{+}\nu\overline{\nu}$
at 5\% level, and
$K_{L} \rightarrow \pi^{0}\nu\overline{\nu}$
at 20\% level (based on the design in the KLEVER project)
~\cite{kaonWhitePaper}.

In Japan, the KOTO experiment is
searching for the decay $K_{L} \rightarrow \pi^{0}\nu\overline{\nu}$
with the high-intensity proton beam from J-PARC.
In the standard model, the branching ratio of 
$K_{L} \rightarrow \pi^{0}\nu\overline{\nu}$ is predicted 
to be $3\times 10^{-11}$.
The KOTO experiment set the current upper limit $3.0\times 10^{-9}$ at 90\% C.L. using the data collected in 2015,
and reached a single event sensitivity of $7.2\times 10^{-10}$ with the data collected in 2016-18.
KOTO will reach a sensitivity 
equivalent to 
the branching ratio of
$\mathcal{O}(10^{-11})$ in the 2020s.

To go beyond the goals of the current KOTO experiment
and explore the effects of new physics at the level of 10\% deviations 
from the standard model prediction,
a new experiment, called KOTO step-2, is proposed
with a new detector and a new beam line
~\cite{Aoki:2021cqa}.
For $K_{L} \rightarrow \pi^{0}\nu\overline{\nu}$ in the standard model,
35 events are expected so far with a
signal-to-background ratio of 0.6,
which will give a discovery of the decay.

The new beam line requires an extension of the Hadron Experimental Facility at J-PARC 
with a second production target (T2) in addition to the current one (T1)~\cite{Aoki:2021cqa}.
In order to use the T1 and T2 simultaneously for also other experiments, 
the beam power of 150~kW is desirable in the slow extraction at J-PARC Main Ring (MR).

J-PARC MR has achieved 64~kW beam power in its slow extraction mode 
with an extraction efficiency of 99.5\%.  
To increase the beam power, the MR is to be operated in faster cycle and 
circulating protons are increased by suppressing the beam instability.  
To aim over 100~kW, further suppression of the beam instability and 
higher extraction efficiency are necessary.  
A new scheme, setting a diffuser in front of the extraction system, 
has successfully demonstrated increasing the extraction efficiency.

\subsection{Muon}

{\flushleft \underline{Charged Lepton Flavor Violation}}\\                     
Discovery of charged lepton flavor violation would be a clear signature
of new physics and such experiments should be advanced intensively.
The MEG experiment set an upper limit on the branching ratio
of the $\mu \rightarrow e\gamma$ process at
$4.2 \times 10^{-13}$,
which is 30 times more stringent than the previous limit from MEGA experiment.
An upgraded experiment, MEG I\hspace{-.1em}I aims for ten-times better sensitivity
with improved detectors and the increased beam intensity.
The upgraded detectors have been successfully commissioned during the engineering run in 2021.
The physics data taking will start from summer 2022.
The projected sensitivity for three DAQ years is $6 \times 10^{-14}$.

Preparations for experiments searching for $\mu - e$ conversion
at a sensitivity of $10^{-14} \sim 10^{-17}$ are underway both in Japan and in the U.S.
The DeeMe experiment at the MLF facility in J-PARC is aiming at
the early realization of $\mathcal{O}(10^{-14})$ sensitivities.
The construction of new beamline (H-line) was completed in January 2022.
The COMET experiment at J-PARC and the Mu2e experiment at Fermilab are now
in the construction phase and they are aiming for sensitivities of $\mathcal{O}(10^{-17})$.
The COMET is one of the flagship experiments of J-PARC employing a staging approach.
The first phase of the experiment (Phase-I), which is aiming for a sensitivity of $\mathcal{O}(10^{-15})$,
 is under construction with a priority defined by the KEK-PIP 2016.
The COMET will realize a high-intensity pulsed proton beam with an extremely small extinction factor 
in addition to large-acceptance pion capture and transport solenoids for the first time.
  Prior to Phase-I, the commissioning of the proton beam and the muon transport
  with a low-intensity beam (Phase-$\alpha$) is under preparation.

{\flushleft \underline{muon $g-2$ and EDM}}\\
The Fermilab $g-2$ experiment released the first results in April 2021 and confirmed
the BNL-E821 result. The combined results give 4.2$\sigma$ deviation from the standard model prediction. 
The improvements in the precision of the standard model prediction are foreseen by
community efforts under the muon $g-2$ theory initiative.
The J-PARC muon $g-2/EDM$ experiment~\cite{Abe:2019thb} will measure $g-2$ 
and the electric dipole moment simultaneously.
It employs a completely new method by using a low-emittance positive muon beam
from a thermal-energy muon source accelerated by linear accelerators.
The experiment is possible to achieve a precision of 0.4 ppm with the
current technical achievements,
and the project is aiming at a final precision of 0.1 ppm.
Construction of the frontend muon beamline (H-line) has been completed.
The first beam of surface muon was delivered to H-line
in January 2022. Further extension of the beamline and construction of muon source and LINAC are
currently in progress.

\subsection{Neutron}

{\flushleft \underline{Neutron EDM}}\\   
The neutron electric dipole moment (EDM) is sensitive to new physics,
since the SM prediction is highly suppressed to be $10^{-32}\,e\cdot$cm (without strong CP-violation).
An upper limit on the neutron EDM of $|d_{n}| < 1.8 \times10^{-26}\,e\cdot\mathrm{cm}$ (90\% C.L.)
was obtained at PSI using ultra-cold neutrons (UCNs) produced with solid deuterium converter.
The intensity of the UCNs was less than the design value due to technical problems.
They are now working on the upgrade of the UCN source and 
the measurement apparatus. 
Developments of high-density UCN sources to study 
the neutron EDM of 10$^{-27}$ to $10^{-28}\,e\cdot $cm is
becoming a ground for international competition. 
At TRIUMF, 
TUCAN collaboration successfully produced the UCNs 
by using a superfluid helium converter in 2017~\cite{TUCAN:2018vmr}.
The UCN source has been upgraded to have higher UCN density by
Japanese institutes KEK and RCNP in charge of the development,
and is now under commissioning.
The EDM measurement apparatus is also being developed in Japan.

{\flushleft \underline{Neutron lifetime}}\\
The neutron lifetime is an important parameter in big bang nucleosynthesis and 
can be an independent systematic test for $V_{ud}$ determination.
In 2021, the US group reported a new value of neutron lifetime in a precision of 0.34 s by using UCNs, 
however, this did not solve the ``neutron lifetime puzzle'', 8.5-s deviation in the lifetime between two methods:
``bottle'' method counting stored UCNs and ``beam'' method counting protons from the neutron decay.
At J-PARC, the lifetime has been measured using a pulsed neutron beam
with a third method counting electrons from the neutron decay.
The method was demonstrated in the first result  published in 2021~\cite{10.1093/ptep/ptaa169}, 
although the accuracy was not enough.
Upgrade and commissioning of both of beamline and apparatus will
improve the accuracy better than 1 s to solve the puzzle in the near future. 

{\flushleft \underline{T-violation in neutron reaction with compound nuclei}}\\
A large enhancement of T-violation in compound nuclei has been theoretically suggested. 
An international collaboration NOPTREX 
plans to search for such T-violation in resonant neutron-nuclei capture reactions.
One candidate of the nuclei is $^{139}$La,
because detailed studies with pulsed neutrons in J-PARC have shown that the violation could be enhanced 
by a factor of $10^6$~\cite{Okudaira:2017dun,Yamamoto:2020xtz}.
Polarizations for both nuclear target and neutron beam are necessary, 
and the techniques of the polarization are being developed.

{\flushleft \underline{Exotic force in neutron interaction}}\\
Neutron interaction is sensitive to exotic interactions mediated by a massive boson which couples to mass,
because deviation from the gravity low is possible in the short range and the neutron is massive neutral particle.
New limits of the Yukawa-type interaction in the range of nanometer were obtained 
by using neutron scattering~\cite{Haddock:2017wav} and interference~\cite{Heacock:2021btd}.

\newpage

\section{Non-accelerator Particle Physics}
\subsection{Particle Physics in Underground Laboratories}
\label{sec:particle_underground}

{\flushleft \underline{Dark Matter Searches}}\\
XMASS-I, a single phase liquid xenon detector, had been taking data over five years 
looking for WIMPs and other new physics and completed its data taking
in 2019. XMASS-I provided results on WIMPs, axions, dark photons, exotic properties 
of neutrinos, a rare decay of a xenon nucleus and possible 
coincidence events with gravitational wave events.
Some of the non-WIMP searches in liquid xenon detectors were pioneered 
and a broader range of dark matter candidates started to be sought by
utilizing the low background of electronic recoils in XMASS-I.
At present final results on WIMPs are being prepared.

In this community, experimental searches for a wide range of WIMP masses
have been conducted with various approaches.
For the heavy WIMPs, the final result from the XENON1T 
experiment excluded WIMP models with a nucleon interaction cross section 
greater than $4.1\times 10^{-47}$cm$^2$  at 30 GeV/c$^2$
and more recently PandaX-4T improved it ($3.8\times 10^{-47}$cm$^2$  at 40 GeV/c$^2$).
XENON1T is also providing the best constraints at lighter WIMPs down to about 
85 MeV by looking for electronic recoils. No significant excess has been
observed in the WIMP searches.

Based on the future plan described in 
the document in 2017~\cite{report2017},
members of the XMASS group started to contribute to the worldwide efforts 
on future WIMP search programs by joining the XENON experiment.
In 2021, XENONnT started to take data. It is expected to provide a sensitivity
down to $1.4\times 10^{-48}$cm$^2$  at 50 GeV/c$^2$ with an exposure of 20 t yr~\cite{XENON:2020kmp}.

An excess observed at low energies of XENON1T is consistent with 
a solar axion signal, 
a solar neutrino signal from an enhanced neutrino magnetic moment, 
and a bosonic dark matter signal with a particle mass of 2.3 keV$/c^2$. 
However, the solar axion and neutrino hypotheses are in strong tension with stellar constraints.  
A tritium background in the detector is also possible and 
XENON1T is unable to confirm or exclude the presence of tritium at this time. 
XENONnT is expected to provide new information on this topic soon.

A part of the XENON members in Japan joined DARWIN~\cite{DARWIN:2016hyl} collaboration in 2020. 
The DARWIN collaboration aims at building the ultimate, 
liquid xenon-based direct detection dark matter detector, 
with a dark matter sensitivity of $10^{-49}$cm$^2$ for the spin-independent case, 
limited by irreducible neutrino backgrounds. 
The detector will have a 40 ton liquid xenon target operated as a dual-phase time projection chamber. 
R\&D of the detector and the experimental site study are underway, and 
its conceptual design report will be available within a few years. 
The members in Japan have been developing a hermetic TPC design~\cite{10.1093/ptep/ptaa141} 
to reduce background originating from radon, and 
photosensors based on a legacy of XMASS low-background technology~\cite{Abe_2020, Ozaki_2021}.
In 2021, groups in Japan have joined in signing a Memorandum of Understanding between 
members of the XENON/DARWIN and LUX-ZEPLIN Collaborations 
towards building this Next-Generation Liquid Xenon Experiment together.

R\&D has been done for production of radio-pure sodium iodide crystals and the
construction of a detector with 150 kg of NaI(Tl), PICOLON,
has been approved.
This is expected to check the result of the DAMA/LIBRA experiment.

As for directional dark matter measurements, 
NEWAGE started to join a worldwide collaborative TPC development framework, CYGNUS, 
and is one of its core groups now. 
Further development of this international cooperation is expected.

An experiment based on cutting-edge technology developed for interferometers in
gravitational wave detection was proposed to search 
for light dark matter axions. This is expected to explore a new area for
parameter space for axions and new gauge bosons.

{\flushleft \underline{$0\nu\beta\beta$ decay}}\\
One of the most important advances in $0\nu\beta\beta$ decay experiments 
since June 2017 is the recent result from KamLAND-Zen 800. 
The experiment started in January 2019 after the installation of a larger inner balloon, 
which houses $^{136}$Xe dissolved in liquid scintillator to increase the amount of this isotope to 745 kg.
In March 2022, the collaboration reported that no signal was observed and 
imposed a lower limit for the $0\nu\beta\beta$ half life of $^{136}$Xe,
$T_{1/2}^{0\nu}> 2.0\times 10^{26}$ yr (90\% C.L.)~\cite{KamLAND-Zen:2022tow}. 
The limit combined with the previous experimental limit by KamLAND-Zen 400 
($T_{1/2}^{0\nu}> 1.07 \times 10^{26}$ yr) is 
$T_{1/2}^{0\nu}> 2.3 \times 10^{26}$ yr (90\% C.L.), 
which is the most stringent limit in the world. 
This result corresponds to $\braket{m_{\beta\beta}}<36$–156 meV and this means that 
we entered the inverse hierarchy region. 
The KamLAND-Zen 800 continues physics run, and further improvement is expected.

To probe down into the inverse hierarchy region, the demand for detectors with high energy resolution 
is increasing because the $2\nu\beta\beta$ events are an unavoidable background. 
The KamLAND group is proceeding with a project targeting $m_{\beta\beta}$=20 meV, 
KamLAND2-Zen, where the energy resolution is improved to about 2\%. 
In the CANDLES experiment, a $0\nu\beta\beta$ decay experiment using $^{48}$Ca, 
collaborative work started with a Korean group for intensive R\&D on a scintillating bolometer, 
which will provide higher energy resolution and good particle discrimination.

Experiments using different nuclei and different methods are important in future $0\nu\beta\beta$ experiments, 
andnew experimental ideas are being developed in Japan such as AXEL and ZICOS.

\subsection{Cosmological Observations}

{\flushleft \underline{Inflation Physics}}\\
CMB polarization experiments provide information about physics beyond the Standard Model, 
such as the inflation physics.
The current best upper limit on the tensor-to-scalar ratio comes from the combined analysis 
using the BICEP/Keck and Planck data, as $r < 0.032$ (95\% C.L.).

The successful POLARBEAR experiment leads to an upgrade project called as Simons Array (SA), 
which contains three telescopes and corresponding receivers. 
The development of the first SA receiver, PB-2A, has been led by Japanese institutions 
including KEK and Kavli IPMU, and deployed in 2018. 
The expected sensitivity of SA after the three years of observations is $\sigma(r) = 0.006$ (68\% C.L.).
 
A novel observational concept is proposed as GroundBIRD that plans to observe a large angular scale 
by a scan strategy with continuous rotation in azimuth from the ground. 
This is the first Japanese-born CMB polarization experiment that reached to the first light in 2019 
and currently observing at Tenerife, Spain.
 
Two US-led and yet Japanese participating projects are called Simons Observatory (SO) and CMB Stage-IV (CMB-S4). 
SO is the upcoming ground telescopes in Atacama, Chile.
The telescopes are scheduled to start science observations in 2023.
Japanese team, utilizing its unique expertise, has been providing key telescope components 
such as the 1-K optics tube, cryogenic half-wave plate, and calibrators.
The expected baseline sensitivity is $\sigma(r) = 0.003$~\cite{Ade_2019}.
CMB-S4 is a proposed ground telescope array by unifying efforts 
in the ground-based CMB experiment community with a prospective deployment in late 2020s.
One of the major science goals is to set an upper limit of $r < 0.001$.

Space-borne CMB missions, WMAP and Planck, have played a key role to probe the inflationary Universe. 
In May 2019, LiteBIRD was selected as the second L-class mission of 
the Institute of Space and Astronautical Science (ISAS) Japan Aerospace Exploration Agency (JAXA).
The prospective launch is in the late 2020s, and currently it is under the conceptual design phase 
with international partners for the design sensitivity of 
$\sigma(r)=0.001$~\cite{litebirdcollaboration2022probing}.
LiteBIRD has been adopted as a priority project of the Science Council Master Plan 2014, 2017, and 2020. 
It has also been selected as one of the plans on the MEXT Road Map 2014, 2017, and 2020.
As seen by the joint constraint on $r$ by the ground telescopes, BICEP/Keck, and Planck today, 
we anticipate the joint analysis using the data from LiteBIRD and CMB-S4 in the future.
The combined data should cover large angular ranges and frequency bands, and thus it opens up 
even deeper sensitivity to the inflationary signal in the 2030s. 

{\flushleft \underline{Dark Energy, Dark Matter and Neutrino Mass}}

The nature of dark energy is one of the most tantalizing problems in cosmology.
The upcoming Subaru Prime Focus Spectrograph (PFS) cosmology program~\cite{2014PASJ...66R...1T}, 
which has been led by Kavli IPMU and  will start in 2024, allows us to measure 
the Hubble expansion rate $H(z)$ and the angular diameter distance $D_A(z)$ 
to $3\%$ fractional accuracies in each of 7 redshift bins over $0.6<z<2.4$, 
which can be used to explore the nature of dark energy over the wide redshift range. 
The ``early'' dark energy model
can be a solution to the $H_0$ tension, and the PFS survey can address whether this is the case. 
The same PFS dataset allows us to robustly measure redshift-space distortion (RSD) effects 
in the distribution of galaxies.
The RSD measurements can be used to not only constrain the strength of gravitational field 
(matter clustering) in large-scale structure, but also make a stringent test of gravity theory 
on cosmological scales, which can be used to address whether modification of gravity is 
an explanation of the cosmic acceleration without the need of introducing mysterious dark energy. 

The nature of dark matter is another unresolved mystery.
Primordial black holes (PBH), which could form in the early universe, are a viable candidate of dark matter. 
Recent observational results have triggered various studies including theoretical studies 
that build a model for PBH formation in the early universe such as inflationary scenario. 
Any observational confirmation of PBH promises transformative progress of our understanding of 
the early universe physics. 
The ultimate microlensing surveys by the Subaru Hyper Suprime-Cam and other telescopes 
in the coming decade will provide us with a conclusive answer to the PBH existence.

Some dwarf galaxies in the Milky Way exhibit a ``core'' in the central region of their mass profile.
The cored profile can be a hint of ultralight dark matter  of $10^{-21}\,{\rm eV}$ mass scales, such as 
axion-like particles inspired by string theory.
The unique wide-field and massively-multiplexed spectroscopic capability of  PFS
enables high-precision reconstruction of dark matter distribution for each of the representative dwarf galaxies.
The PFS measurements of dwarf galaxies 
can give a robust test of ``wave'' or ``fuzzy'' dark matter scenario. 

Neutrino mass takes part in the role of dark matter in structure formation due to its weak interaction nature. 
The expected sensitivity of SA after the three years of observations is 
$\sigma(\sum{m_\nu})$= 40~meV (68\% C.L.) for the sum of neutrino masses 
by combining the DESI results of baryon acoustic oscillations.
The Subaru PFS and US-led DESI surveys are both designed to independently achieve a precision of 
$\sigma(\sum m_\nu)\simeq 20\,{\rm meV}$ 
until around 2025.
Hence these cosmology programs will enable a $5\sigma$ detection of $\sum m_\nu$, rather than the upper limit, 
if the neutrinos follow the inverted mass hierarchy that predicts the lower limit of 
$\sum m_\nu\simeq 100\,{\rm meV}$. 
Any claim of the neutrino mass detection will need 
independent confirmation, which is what PFS and DESI are expected to attain.
The expected precision of $\sigma(\sum m_\nu)\simeq 20\,{\rm meV}$ is currently limited 
by the uncertainty in the optical depth $\tau$ from the Planck data.
The high-precision $\tau$ measurements by upcoming CMB polarization experiments probing large angular scales, 
including GroundBIRD and eventually cosmic variance limited measurements by LiteBIRD, 
will further improve the determination of the neutrino mass, 
such as $\sigma(\sum m_\nu)\simeq 15\,{\rm meV}$ after 2030.

The upcoming ultimate panoramic surveys such as the Vera C. Rubin Observatory's LSST (2024), 
the ESA Euclid (2023) and the NASA Roman Space Telescope (2026) promise to advance
our understanding of the Universe, including exploration of the nature of dark matter and 
dark energy with unprecedented precision. 
Even in the coming decade, only the PFS has the wide-field and massively-multiplex spectroscopic capabilities 
among other 8m-class telescopes, and 
is uniquely positioned to play synergetic and complementary roles with these cosmology surveys.

\newpage

\noindent 
{\large\bf Acknowledgements} \\

The editors thank 
J.~Tian, 
T.~Nakamoto, K.~Terashi, M.~Togawa, Y.~Okumura,
K.~Sakashita, M.~Yokoyama, 
M.~Masuzawa, Y.~Suetsugu, M.~Tobiyama, Y.~Ushiroda, Michael Roney, Soeren Prell,
T.~Mibe,  M.~Kitaguchi,
M.~Yamashita, K.~Martens, Y.~Kishimoto,
M.~Takada, T.~Matsumura, K.~Nakayama,
T.~Abe, T.~Higo, and M.~Otani
for helpful inputs.  
\newpage


\bibliographystyle{JHEP} 
\bibliography{ref} 

\end{document}